\documentclass[preprint,showpacs,preprintnumbers,amsmath,amssymb]{revtex4}

\usepackage{graphicx}
\usepackage{dcolumn}
\usepackage{bm}
\usepackage{amsmath}

\newcommand{\qed}{\nobreak \ifvmode \relax \else
      \ifdim\lastskip<1.5em \hskip-\lastskip
      \hskip1.5em plus0em minus0.5em \fi \nobreak
      \vrule height0.75em width0.5em depth0.25em\fi}

\begin{document}

\preprint{}

\title{Two-Qubit Rational-Valued Entanglement-Boundary Probability Densities and a Fisher Information  Equality Conjecture}
\author{Paul B. Slater}%
\email{slater@kitp.ucsb.edu}
\affiliation{%
University of California, Santa Barbara, CA 93106-4030\\
}%
\date{\today}

\begin{abstract}
We consider a pair of  one-parameter ($\alpha$) families of generalized two-qubit determinantal Hilbert-Schmidt probability distributions, $p_{\alpha}(|\rho^{PT}|)$ and 
$q_{\alpha}(|\rho|)$, where $\rho$ is a $4 \times 4$ density matrix, $\rho^{PT}$, its partial transpose, with $|\rho^{PT}| \in [-\frac{1}{16},\frac{1}{256}]$ and $|\rho| \in [0, \frac{1}{256}]$. The Dyson-index-like (random matrix) parameter $\alpha$ is $\frac{1}{2}$  
for the 9-dimensional generic two-rebit systems,  1 for the 15-dimensional generic two-qubit systems,$\ldots$. 
Numerical (moment-based probability-distribution-reconstruction) analyses suggest the conjecture that the Fisher information--a measure over $\alpha$--is identical for the two distinct families. 
Further, we study the values of $p_{\alpha}(0)$, the probability densities at the separability-entanglement boundary, with evidence strongly indicating that $p_2(0) = \frac{7425}{34}= \frac{3^2 \cdot 5^2 \cdot 11}{2 \cdot 17}$, $p_3(0)= \frac{7696}{69} = \frac{2^4 \cdot 13 \cdot 17}{3 \cdot 23}$, $p_4(0) = 
\frac{14423395}{266104} =\frac{5 \cdot 7^2 \cdot 17 \cdot 3463}{2^3  \cdot 29  \cdot 31  \cdot 37},\ldots$. Despite extensive results of such a nature, we have not yet succeeded--in contrast with the corresponding rational-valued separability probabilities (http://arxiv.org/abs/1301.6617)--in generating an underlying, explanatory ("concise") formula for $p_{\alpha}(0)$, even though  the corresponding denominators of both sets of rational-valued results are  closely related, 
having almost identical (small) prime factors. The first derivatives $p_{\alpha}^{'}(0)$ are positive for 
$\alpha = \frac{1}{2}$ and 1, but negative for $\alpha  > 1$.
\end{abstract}


\maketitle
\section{Introduction}
The nature of the boundary between separable and entangled quantum states is a subject of interest in the study of the "geometry of entanglement sudden death"
\cite[Fig. 1(ii)]{Cunha}. Our analyses will provide information concerning the density of states in the vicinity of this boundary. Additionally, they suggest a conjecture pertaining to the Fisher information of two pertinent one-parameter families of determinantal probability distributions. Let us now place our study, and the accompanying results, in the context from which they   emerged.

{\.Z}yczkowski, Horodecki, Sanpera and Lewenstein, in a much cited 1998 article \cite{ZHSL},  posed "the question of how many entangled or, respectively, separable states there are in the set of all quantum states". In \cite{slaterJModPhys}, we, in essence, provided a "concise" and highly generalized answer--though one still lacking  a formal, rigorous demonstation--to this fundamental question in the context of two-qubit systems endowed with Hilbert-Schmidt (Euclidean/flat) measure \cite{szHS,ingemarkarol}. The formula--for the Hilbert-Schmidt probability $P(\alpha)$ that a generic $2 \times 2$ quantum system is separable--that was reported there took the form 
\begin{equation} \label{Hou1}
P(\alpha) =\Sigma_{i=0}^\infty f(\alpha+i),
\end{equation}
where
\begin{equation} \label{Hou2}
f(\alpha) = P(\alpha)-P(\alpha +1) = \frac{ r(\alpha) 2^{-4 \alpha -6} \Gamma{(3 \alpha +\frac{5}{2})} \Gamma{(5 \alpha +2})}{3 \Gamma{(\alpha +1)} \Gamma{(2 \alpha +3)} 
\Gamma{(5 \alpha +\frac{13}{2})}},
\end{equation}
and
\begin{equation} \label{Hou3}
r(\alpha) = 185000 \alpha ^5+779750 \alpha ^4+1289125 \alpha ^3+1042015 \alpha ^2+410694 \alpha +63000 = 
\end{equation}
\begin{displaymath}
\alpha  \bigg(5 \alpha  \Big(25 \alpha  \big(2 \alpha  (740 \alpha
   +3119)+10313\big)+208403\Big)+410694\bigg)+63000.
\end{displaymath}
Here $\alpha$ is a Dyson-index-like (random matrix) parameter that takes the value 
$\frac{1}{2}$ for the 9-dimensional two-rebit systems, 1 for the 15-dimensional two-qubit systems, and 2 for the 27-dimensional 
two-quater(nionic)bit systems,\ldots. (We have the relation $ n =3 +12 \alpha$, where $n$ is the dimensionality of the associated generic generalized $4 \times 4$ density matrices.)

These separability probabilities were computed--in accordance with the seminal Peres-Horodecki results \cite{asher,michal}--as the cumulative probabilities over the nonnegative interval 
$|\rho^{PT}| \in [0, \frac{1}{256}]$ of probability distributions 
$p_{\alpha}(|\rho^{PT}|)$, with $|\rho^{PT}| \in [-\frac{1}{16},\frac{1}{256}]$. For $\alpha= \frac{1}{2}$, 1 and 2, the formula yields (to arbitrarily high precision)
$P(\alpha) = \frac{29}{64} \approx 0.453125, \frac{8}{33} \approx 0.242424$ (cf. \cite[eq. (B7)]{joynt} \cite[sec. VII]{Fonseca-Romero}) and $\frac{26}{323} \approx 0.0804954$, respectively. In the calculations of these quantities, 
the moment formula 
\cite[sec. D.4]{MomentBased} 
\begin{equation} \label{nequalzero}
\left\langle \left\vert \rho^{PT}\right\vert ^{n}\right\rangle =\frac
{n!\left(  \alpha+1\right)  _{n}\left(  2\alpha+1\right)  _{n}}{2^{6n}\left(
3\alpha+\frac{3}{2}\right)  _{n}\left(  6\alpha+\frac{5}{2}\right)  _{2n}} +
\end{equation}
\begin{displaymath}
\frac{\left(  -2n-1-5\alpha\right)  _{n}\left(  \alpha\right)  _{n}\left(
\alpha+\frac{1}{2}\right)  _{n}}{2^{4n}\left(  3\alpha+\frac{3}{2}\right)
_{n}\left(  6\alpha+\frac{5}{2}\right)  _{2n}}~_{5}F_{4}\left(
\genfrac{}{}{0pt}{}{-\frac{n-2}{2},-\frac{n-1}{2},-n,\alpha+1,2\alpha
+1}{1-n,n+2+5\alpha,1-n-\alpha,\frac{1}{2}-n-\alpha}%
;1\right) 
\end{displaymath}
was employed in  a (Mathematica-implemented) Legendre-polynomial-based probability-distribution reconstruction procedure of Provost \cite{Provost} (cf. \cite{cohentan}).

\begin{figure}
\includegraphics{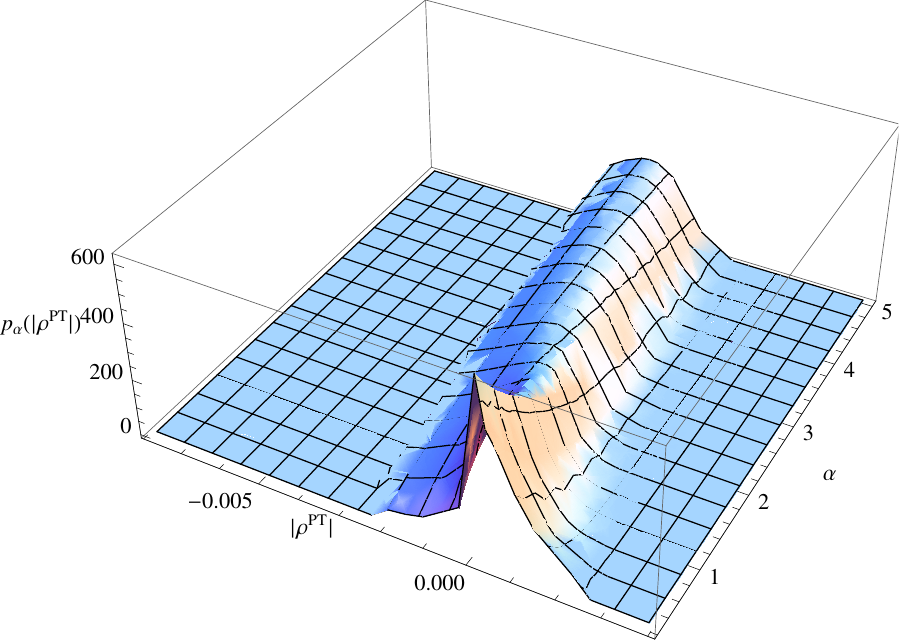}
\caption{\label{fig:AlphaFamilyPlot}The $\alpha$-parameterized family of probability distributions $p_{\alpha}(|\rho^{PT}|)$ over $[-\frac{1}{16},\frac{1}{256}]$ estimated on the basis of the first 245 Hilbert-Schmidt $\alpha$-specific moments  of $|\rho^{PT}|$, given by (\ref{nequalzero})}
\end{figure}

Other than the separability probabilities given by the concise formula 
(\ref{Hou1})-(\ref{Hou3}) for each value of $\alpha$, however, nothing additional appears to be specifically known about the $\alpha$-parameterized family of probability distributions over $[-\frac{1}{16},\frac{1}{256}]$, which we denote as $p_{\alpha}(|\rho^{PT}|)$. (Fig.~\ref{fig:AlphaFamilyPlot} is a plot of this family, estimated on the basis of the first 245 Hilbert-Schmidt  moments (\ref{nequalzero}) of $|\rho^{PT}|$.)

We have, thus, engaged in a research program intended to increase our knowledge of these probability distributions, which are of clear major quantum-information-theoretic interest. Our long-range goal is to fully characterize them--that is, develop explicit formulas, if possible. From such formulas, one should then be able to derive/confirm the already-reported separability probabilities \cite{slaterJModPhys} as the cumulative probabilities of $p_{\alpha}(|\rho^{PT}|)$ over $|\rho^{PT}| \in [0, \frac{1}{256}]$.

Let us observe that twice the fourth root of the absolute value of the determinant $|\rho^{PT}|$ is a two-qubit "monotone under pure local operations preserving dimensions and classical communication and provides tight upper and lower bounds on concurrence" 
\cite[eq. (5)]{Demianowicz}. (If $|\rho^{PT}| \geq 0$, this "determinant-based measure" is taken to be zero.)
\section{Probability density estimates at $|\rho^{PT}|=0$}
Possibly the most natural additional question to be first addressed is what is the value of the "$y$-intercept" of the probability distribution--that is the value the curve/function attains at $|\rho^{PT}|=0$ (the separability-entanglement boundary). Such analyses would appear to be of relevance--among other areas--in studying the geometry of  entanglement sudden death \cite{Cunha,Cunha2}.

\begin{figure}
\includegraphics{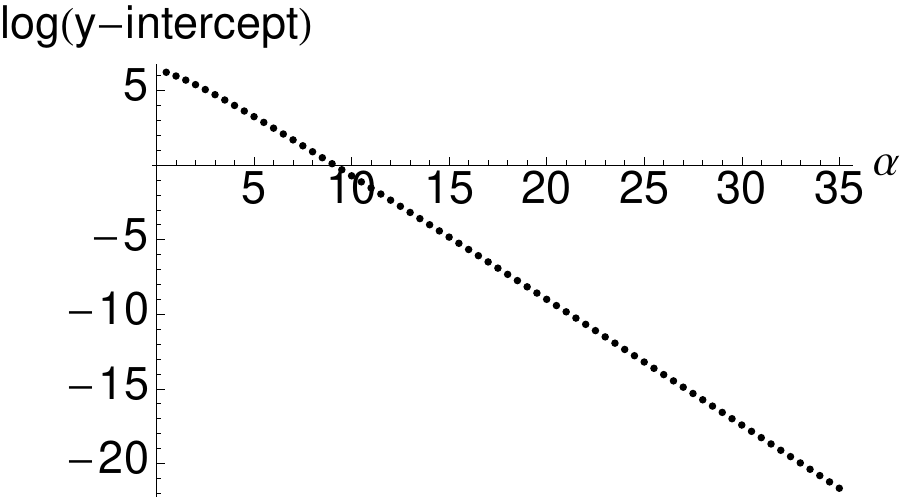}
\caption{\label{fig:DerivativePlotUnbalanced}Logarithms of $y$-intercepts $p_{\alpha}(0)$ of estimated (based on 3,000 moments (\ref{nequalzero})) Hilbert-Schmidt generalized two-qubit probability distributions over $|\rho^{PT}| \in [-\frac{1}{16},\frac{1}{256}]$}
\end{figure}

To begin, we estimated the $y$-intercepts jointly for the seventy integral and half-integral values $\alpha= 
\frac{1}{2}, 1, \frac{3}{2},\ldots 35$, using the corresponding first 3,000 moments of $|\rho^{PT}|$ \cite[sec. D.4]{MomentBased}--given by 
(\ref{nequalzero})--in the Legendre-polynomial probability-distribution reconstruction procedure \cite{Provost}. The logs of these seventy estimates are displayed in Fig.~\ref{fig:DerivativePlotUnbalanced}.  (A least-squares fit to  this highly linear plot is provided by $7.35633 -0.823151 \alpha$.)

\begin{figure}
\includegraphics{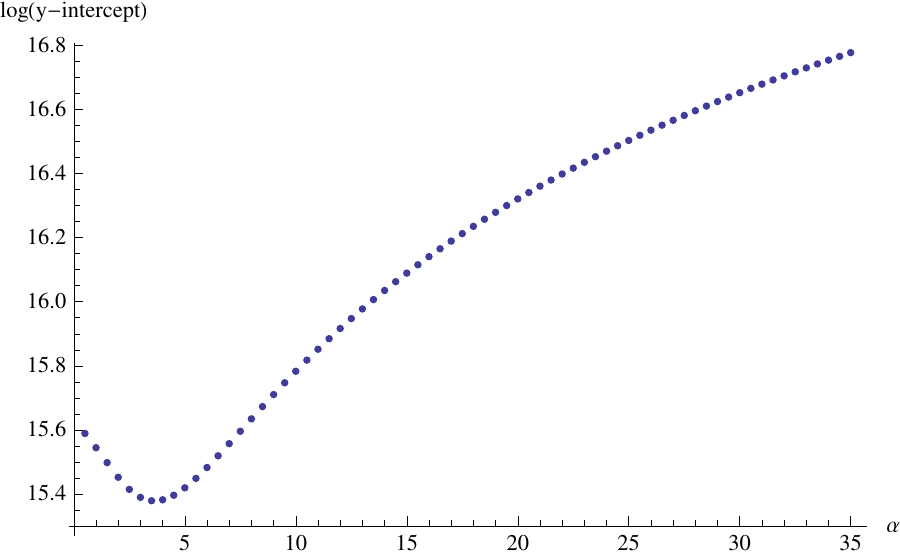}
\caption{\label{fig:DerivativePlotBalanced}Logarithms of $y$-intercepts of estimated (based on 2,600 moments (\ref{BalancedMoments})) Hilbert-Schmidt generalized two-qubit probability distributions over 
$|\rho| |\rho^{PT}| \in [-\frac{1}{110592},\frac{1}{256^2}]$}
\end{figure}

Further, in Fig.~\ref{fig:DerivativePlotBalanced}, we show the analogous plot (exhibiting interesting non-monotonic behavior), based now on the estimated probability distribution of $|\rho| |\rho^{PT}|$ over the interval 
$[-\frac{1}{110592}, \frac{1}{256^2}]$--rather than $|\rho^{PT}|$ over $[-\frac{1}{16}, \frac{1}{256}]$. (We note that $\frac{1}{110592} = 2^{-12} \cdot 3^{-3}$.) Here 2,600 moments \cite[sec. D.4]{MomentBased},
\begin{equation} \label{BalancedMoments}
\left\langle \left\vert \rho\right\vert ^{n}\left\vert \rho^{PT}\right\vert
^{n}\right\rangle 
=\frac{\left(  2n\right)  !\left(  1+\alpha\right)  _{2n}\left(
1+2\alpha\right)  _{2n}}{2^{12n}\left(  3\alpha+\frac{3}{2}\right)
_{2n}\left(  6\alpha+\frac{5}{2}\right)  _{4n}}~_{4}F_{3}\left(
\genfrac{}{}{0pt}{}{\ -n,\alpha,\alpha+\frac{1}{2},-4n-1-5\alpha
}{-2n-\alpha,-2n-2\alpha,\frac{1}{2}-n}%
;1\right)  .
\end{equation}
were utilized in the reconstruction procedure. Rates of convergence are much slower in this "balanced" moment case than in the previous "unbalanced" case, so our further analyses here will fully focus on the unbalanced instance.

In Table~\ref{table:Table1} we give some of our early estimates for the $y$-intercepts ($p_{\alpha}(0)$) as a function of $\alpha$--and, in some cases, advance apparently exact rational values for them. We are strongly confident--based on still further systematic computations with greater than 10,000 moments--in knowing the exact values for $\alpha=2,\ldots, 46$. Nevertheless, Mathematica has not yielded an explanatory formula--as it accomplished in the separability-probability analysis with $\alpha=1,\ldots,28$. Interestingly, most of the prime factors of the denominators of these exact results are the same as the prime factors of the already-reported corresponding exact separability probabilities. By way of illustration, the denominator of $p_{12}(0)$  
(Table~\ref{table:Table1}) is 
$3^2 \cdot 79 \cdot 83 \cdot 89 \cdot 97 \cdot 101 \cdot 103 \cdot 107 \cdot 109 \cdot 113$, while that of the separability probability for 
$\alpha=12$ is identical to this, but for the replacement of $3^2$ by 
$3 \cdot 11^2$. The largest prime factors of $p_{\alpha}(0)$ never exceed those of the corresponding separability probabilities, and can, on occasion, be less.

Rates of convergence markedly increase as $\alpha$ itself increases, being weakest, unfortunately, in  two  (two-rebit and two-qubit) instances of immediate interest, $\alpha =\frac{1}{2}$ and 1. Also, estimates converge monotonically upward for $\alpha < 10$, as the number of moments inputted is increased, so, it is clear, that the  corresponding reported results provide {\it lower} bounds on true values. (It has been suggested that the use of a Jabobi-polynomial-based  probability-distribution reconstruction procedure \cite[p. 742]{Provost} might exhibit superior convergence properties in these cases ($\alpha = \frac{1}{2}, 1$) than the Legendre-polynomial-based procedure we have so far employed. However, our attempts to pursue such a course
seemed to yield rather unstable estimates. The Legendre-based procedure starts with a uniform baseline density, while the Jacobi-based method commences with a beta distribution fitted to the first two moments.)
\begin{table}
\caption{$y$-intercept Estimates $p_{\alpha}(0)$}
\centering
\begin{tabular}{c c c c}
\hline\hline
$\alpha$ & \# moments & Estimated $y$-intercept\footnote{Estimates are reported to the number of decimal places that remain unchanged even after the incorporation of the last one hundred moments employed. For $\alpha < 11$, the estimates converge monotonically upward--providing lower bounds on true values--an observation we take into account in advancing the candidate values.} & Candidate Exact Value \\ [0.5ex]
\hline
$\frac{1}{2}$  & 12,700 & 503.17 (Fig.~\ref{fig:TwoRebitPlot})& \\ 
1 & 11,445 & 389.99 (Fig.~\ref{fig:TwoQubitPlot})& 390 = $2 \cdot 3 \cdot 5 \cdot 13$ \\
$\frac{3}{2}$ & 10,920 & 296.34924 &\\
2 & 10,500 & 218.3823524 (Fig.~\ref{fig:TwoQuaterbitPlot})& $\frac{7425}{34}= \frac{3^2 \cdot 5^2 \cdot 11}{2 \cdot 17}$ \\
$\frac{5}{2}$ & 8,100 & 157.34353772 & $\frac{11698020}{74347} = \frac{2^2 \cdot 3^5 \cdot 5 \cdot 29 \cdot 83}{7 \cdot 13 \cdot 19 \cdot 43}$ \\
3 & 8,400 & 111.536231883 (Fig.~\ref{fig:Alpha3Plot1})& $\frac{7696}{69} = \frac{2^4 \cdot 13 \cdot 37}{3 \cdot 23}$ \\
$\frac{7}{2}$ & 5,900 & 78.1142995070 & \\
4 & 7,500 & 54.202097676095 & $\frac{14423395}{266104} =\frac{5 \cdot 7^2 \cdot 17 \cdot 3463}{2^3  \cdot 29  \cdot 31  \cdot 37}$\\
$\frac{9}{2}$ & 5,600 & 37.335822413322 &  \\
5 & 7,500 &  25.5666434018285 & $\frac{1959591816}{76646425} = 
\frac{2^3 \cdot 3^2 \cdot 7 \cdot 13 \cdot 61 \cdot 4903}{5^2 \cdot 37 \cdot 41 \cdot 43 \cdot 47}$\\
$\frac{11}{2}$ & 5,900 & 17.422692059774 &\\
6 & 7,000 & 11.82486375905315966 & $\frac{1090539700}{92224293} =\frac{2^2 \cdot 5^2 \cdot 31 \cdot 61 \cdot 73 \cdot 79}{3 \cdot 7 \cdot 41 \cdot 43 \cdot 47 \cdot 53}$ \\
$\frac{13}{2}$ & 5,300 & 7.9980614144240790 & \\
7 & 9,240 & 5.393779773089919955165 & $\frac{1111265181870}{206027169929} =
\frac{2 \cdot 3 \cdot 5 \cdot 17 \cdot 29 \cdot 75136253}{7^3 \cdot 47 \cdot 53 \cdot 59 \cdot 61 \cdot 67}$ \\
8 & 7,600 & 2.4351194168151571075324 & $\frac{5161595531499}{2119647806944}  =\frac{3^2 \cdot 11 \cdot 41 \cdot 97 \cdot 13109713}{2^5 \cdot 53 \cdot 59 \cdot 61 \cdot 67 \cdot 71 \cdot 73}$\\
9 & 7,000 & 1.090697637286080985468617 & $\frac{241329925246960}{221261985904221} =\frac{2^4 \cdot 5 \cdot 11 \cdot 23 \cdot29 \cdot 37 \cdot 97 \cdot 109 \cdot 1051}{3^3 \cdot 59 \cdot 61 \cdot 67 \cdot 71 \cdot 73 \cdot 79 \cdot 83}$\\
10 & 7,000 & 0.485464865631244541283680427 & $\frac{238572701926614}{491431448116025} = \frac{2 \cdot 3^4 \cdot 11^2 \cdot 41 \cdot 1777 \cdot 167051}{5^2 \cdot 67 \cdot 71 \cdot 73 \cdot 79 \cdot 83 \cdot 89 \cdot 97}$ \\
11 & 8,460 & 0.21497802182219253282721028797 & 
$\frac{8495128297888277220}{39516264155200777043} =\frac{2^2 \cdot 3^ \cdot 5 \cdot 7 \cdot19 \cdot 11861033 \cdot 29917361}{11^2 \cdot 71 \cdot 73 \cdot 79 \cdot 83 \cdot 89 \cdot 97 \cdot 101 \cdot 103 \cdot 107}$\\
12 & 7,770 & 0.09479712852190406271015841397642 & 
$\frac{662143216303197685}{6984844653287163753} =\frac{5 \cdot 7^2 \cdot 29 \cdot 61 \cdot 6719 \cdot 227380583}{3^2 \cdot 79 \cdot 83 \cdot 89 \cdot 97 \cdot 101 \cdot 103 \cdot 107 \cdot 109 \cdot 113}$\\
13 & 8,500 & 0.041653632242518789148830886711329662 & $\frac{96610235322575857186}{2319371207775705136339}$\\
14 & 6,325 &  0.01824714034728285015352436495808 & 
$\frac{5306622762208739719059}{290819419438451333002342}$\\
15 & 5,500 & 0.0079726370509246020901593058511607 & 
$\frac{2771311717360093058282}{347602894708307171124675}$\\
16 & 6,100 &0.003475526958451101365798473125180273 & $\frac{6825762937391268808429713}{1963950508510291938204587648}$ \\
\hline
\end{tabular}
\label{table:Table1}
\end{table}

\begin{figure}
\includegraphics{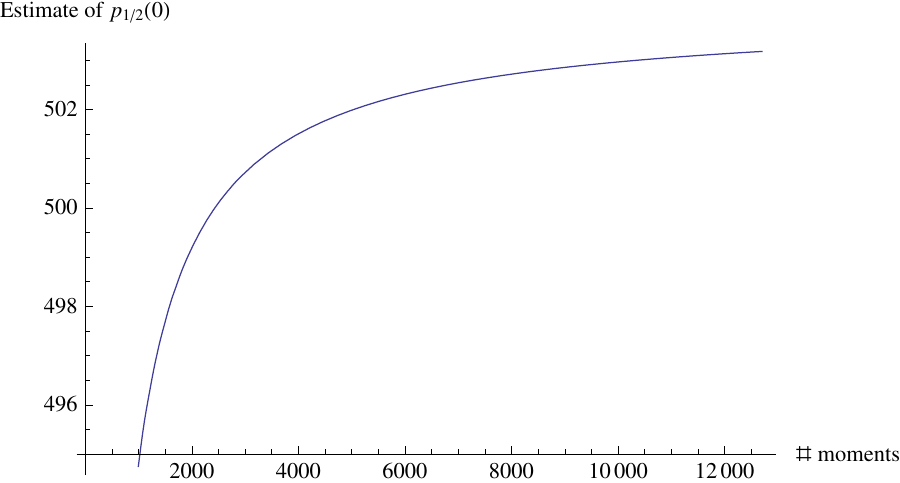}
\caption{\label{fig:TwoRebitPlot}Estimates 
(relatively slowly converging)--as a function of the numbers of moments employed--of the two-rebit probability density 
$p_{\frac{1}{2}}(0)$. At the final (12,700{\it th}) iteration recorded, the estimate is 503.17735, while at the 12,600{\it th} iteration it was slightly less,  503.17064.}
\end{figure}

\begin{figure}
\includegraphics{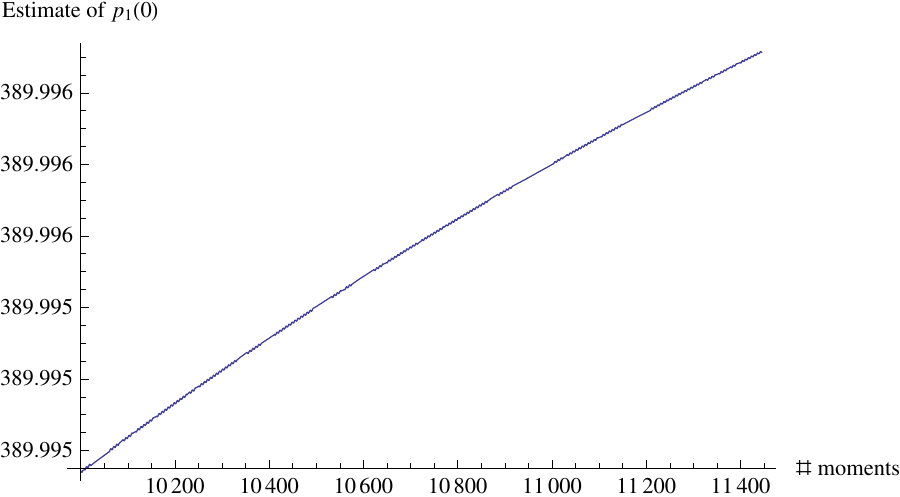}
\caption{\label{fig:TwoQubitPlot}Estimates--as a function of the numbers 
(greater than 10,000) 
of moments employed--of the two-qubit probability density 
$p_{1}(0)$. At the final (11,445{\it th}) iteration recorded, the estimate is 389.99611337, while at the 11,440{\it th} iteration it was slightly less, 389.99611028.}
\end{figure}

\begin{figure}
\includegraphics{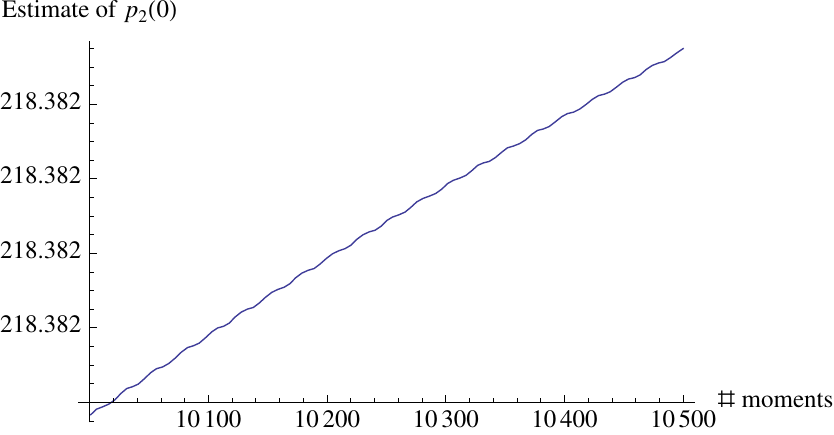}
\caption{\label{fig:TwoQuaterbitPlot}Estimates--as a function of the numbers (greater than 10,000) 
of moments employed--of the two-quaterbit probability density 
$p_{2}(0)$. At the final (10,500{\it th}) iteration recorded, the estimate is 218.382352474952, while at the 10,499{\it th} iteration it was slightly less, 218.382352474705. 
We note that, quite closely, $\frac{7425}{34} \approx 218.38235294$.}
\end{figure}

\begin{figure}
\includegraphics{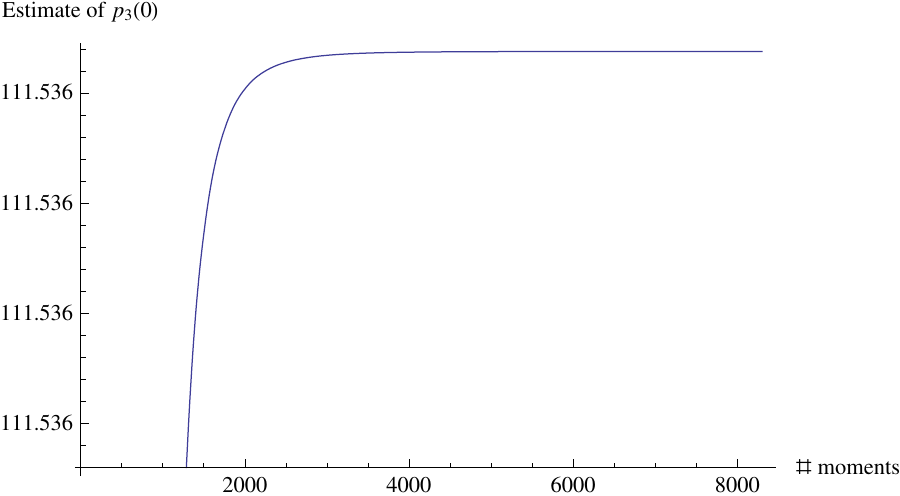}
\caption{\label{fig:Alpha3Plot1}Estimates--as a function of the numbers of moments employed--of the generalized two-qubit probability density 
$p_{3}(0)$. At the final (8,400{\it th}) iteration recorded, the estimate is 111.5362318835796, while at the 8,300{\it th} iteration it was slightly less,  111.5362318835462. Note that, remarkably closely, $\frac{7696}{69} \approx 111.5362318840579$.}
\end{figure}

\begin{figure}
\includegraphics{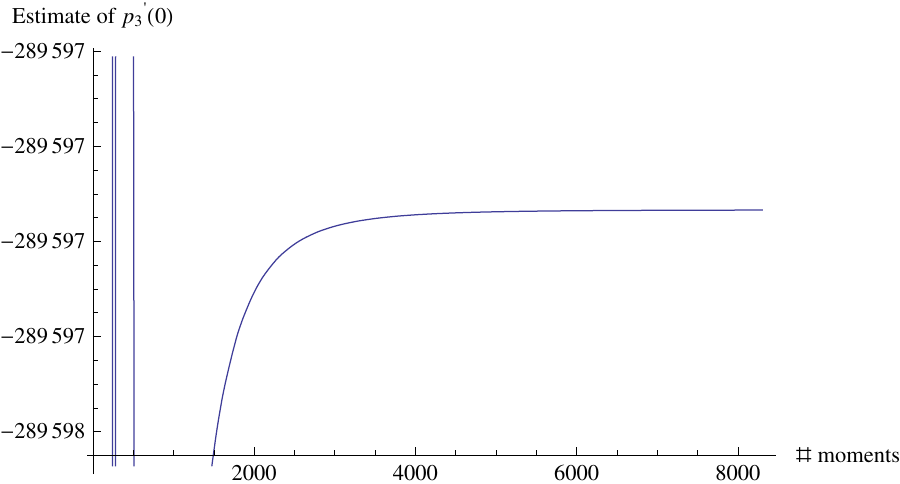}
\caption{\label{fig:Alpha3Plot2}Estimates--as a function of the numbers of moments employed--of the first derivative $p_{3}^{'}(0)$ of the generalized two-qubit probability density for $\alpha= 3$. At the final (8,400{\it th}) iteration recorded, the estimate is -289597.1343229, while at the 
8,300{\it th} iteration it was slightly less, -289597.1343453.}
\end{figure}

In Fig.~\ref{fig:TwoRebitPlot}, we show the estimates of the two-rebit density $p_{\frac{1}{2}}(0)$ as a function of the number of moments employed in the inversion procedure. In Figs.~\ref{fig:TwoQubitPlot} and \ref{fig:TwoQuaterbitPlot}, we show the estimates of the two-qubit density $p_{1}(0)$ and two-quaterbit density $p_{2}(0)$ as a function of the number  of moments exceeding 10,000 employed.
In Fig.~\ref{fig:Alpha3Plot1}, we show the estimates of the generalized two-qubit density $p_{3}(0)$ as a function of the number of moments employed in the inversion procedure, while in Fig.~\ref{fig:Alpha3Plot2}, we show the 
corresponding estimates of the first derivative of the density $p_{3}^{'}(0)$. (In an effort to discern an exact value for at least one $\alpha$-specific value of  $p_{\alpha}^{'}(0)$, for $\alpha=4$, due to the apparent speedy convergence properties there, we employed 
11,480 moments in the Provost procedure \cite{Provost} and obtained an estimate of 
-204971.49610681565 [continually decreasing in absolute value]. The WolframAlpha website suggested a number of possible exact values, 
such as $-\frac{\sqrt{414654790893}}{\pi} \approx -204971.496106763$.)
Of course, if one had in addition to the exact value of $p_{\alpha}(0)$, derivatives of higher order of $p_{\alpha}(|\rho^{PT}|)$ at $|\rho^{PT}|=0$, one could attempt a Taylor series expansion about that distinguished point. However, in practical terms, rates of convergence of estimates--using the Legendre-polynomial reconstruction procedure \cite{Provost}--become increasingly slow as the order of the derivative increases 
(cf. \cite{cohentan}). (Therefore, at the present stage of analysis, we are unable to  advance any possible, plausible candidates for exact values for these derivatives.) 

\begin{figure}
\includegraphics{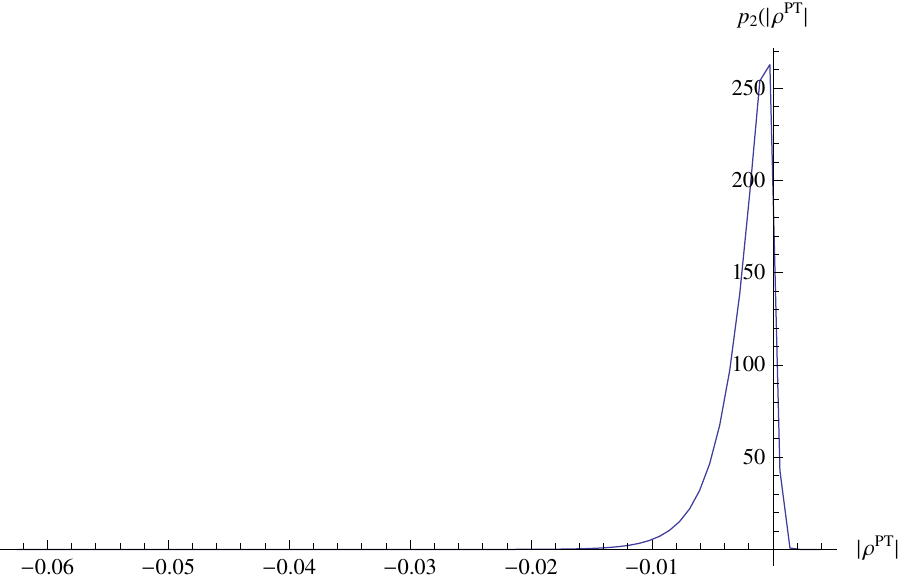}
\caption{\label{fig:TwoQuaterBitPlot2}The two-quaterbit ($\alpha=2$) determinantal probability distribution, based on the first 5,000 moments (\ref{nequalzero}) of 
$|\rho^{PT}|$}
\end{figure}

In Table~\ref{table:Table2}, we list some estimates we have obtained of the first derivative, $p^{'}_{\alpha}(0)$,  for various values of $\alpha$. We do note that these values are all negative (decreasingly so as $\alpha$ increases), with the exception of  
the generic 9-dimensional two-rebit and 15-dimensional two-qubit cases, 
$\alpha =\frac{1}{2}, 1$, respectively.
\begin{table}
\caption{Estimates of the first derivatives  $p^{'}_{\alpha}(0)$}
\centering
\begin{tabular}{c c c}
\hline\hline
$\alpha$ & \# moments & Estimated first derivative\footnote{Digits to the right of the decimal point are reported to the extent they have remained unchanged after the incorporation of the last one hundred moments. If no such digits are given, simply the integral part of the last estimate obtained is presented.}\\ [0.5ex]
\hline
$\frac{1}{2}$ & 12,700 & 6410356.  \\
1 & 10,000 & 339196. \\
$\frac{3}{2}$ & 8,700 & -162646. \\
2 & 10,000 & -289800.  \\
$\frac{5}{2}$ & 8,100 & -311862.10 \\
3 & 8,400 & -289597.134 (Fig.\ref{fig:Alpha3Plot2}) \\
$\frac{7}{2}$ & 5,800 & -249449.43420 \\
4 & 11,480 & -204971.49610681 \\
$\frac{9}{2}$ & 5,600 & -162971.8709833\\
5 & 7,500 & -126442.27058586704\\
$\frac{11}{2}$ & 5,900 & -96249.9203477421 \\
6 & 7,000 & -72156.15107106951 \\
$\frac{13}{2}$  & 5,300 &  -53418.9817136894241\\
7 & 5,300 & -39134.0668213575364 \\
8 & 7,600 & -20472.80618300754317070 \\
9 & 4,800 & -10425.4751799502175604\\
10 & 4,800 & -5199.062697559066885103 \\
11 & 3,500 & -2549.69007308405673430 \\
12 & 5,300 & -1233.4041330231218163946548 \\
13 & 9,500 & -589.88846332521988518644909\\
14 & 4,900 & -279.4104922820162161974541786  \\
15 & 5,500 & -131.2569443034131032177602791621  \\
\hline
\end{tabular}
\label{table:Table2}
\end{table}
Further, since the Legendre-polynomial probability distribution reconstruction algorithm was relatively fast in the two-quaterbit ($\alpha =2$) case, we pursued some additional analyses there, complementing our results of $\frac{7425}{34}$ for $p_{2}(0)$ (Table~\ref{table:Table1}) and 
$\frac{26}{323}$ for the cumulative ("separability") probability over 
$[0,\frac{1}{256}]$. Thus, based on 5,000 moments, we obtained  estimates of $p_{2}(-\frac{1}{256})$ and $p_{2}(\frac{1}{256})$ of 83.56342 and 0.020157, respectively. Also, our estimates of the cumulative probability over the intervals $[-\frac{1}{256},\frac{1}{256}]$ and  $[\frac{1}{512},\frac{1}{256}]$ were 0.81418669739 and 0.000002538812, respectively.
In Fig.~\ref{fig:TwoQuaterBitPlot2} we attempt a companion reconstruction of the (sharply-peaked) two-quaterbit determinantal probability distribution $p_{2}(|\rho^{PT}||)$.
While the separability probabilities for {\it half}-integral $\alpha$ are, 
it seem abundantly clear, rational-valued, along with the ones for 
integral $\alpha$, we presently lack any evidence/argument that 
the values of $p_{\alpha}(0)$ are rational-valued for half-integral
$\alpha$ (if for no other reason than that we have been predominantly
concerned with the integral cases).

\begin{figure}
\includegraphics{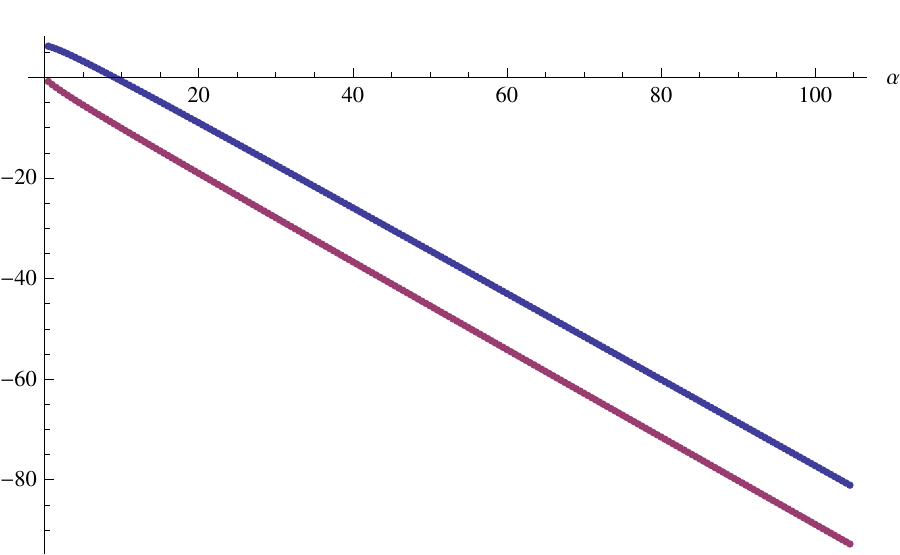}
\caption{\label{fig:JointPlot}Joint plot of the logs of the $\alpha$-specific separability 
probabilities (lower line) together with the logs of the estimated values (upper line), based on 10,700 moments, of the $y$-intercepts
$p_{\alpha}(0)$ for the 210 values $\alpha=\frac{1}{2},1,\frac{3}{2},2,\ldots 105$.}
\end{figure}

\begin{figure}
\includegraphics{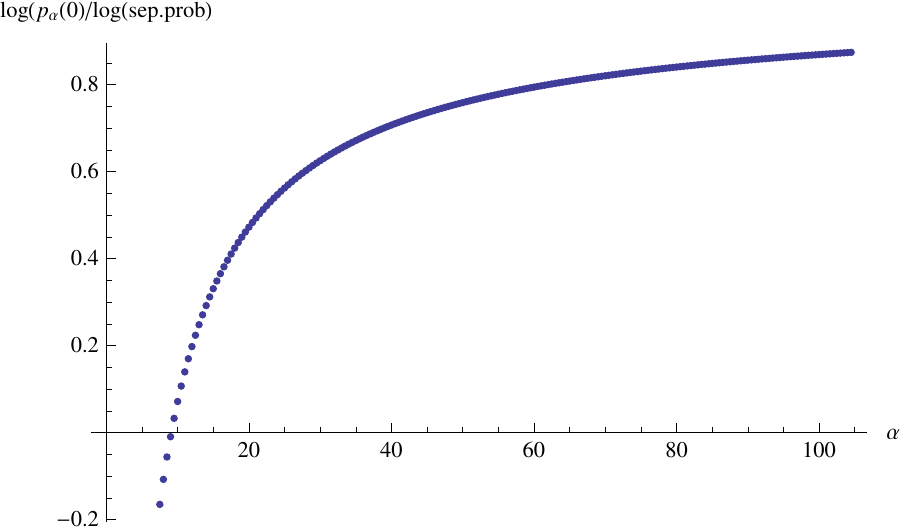}
\caption{\label{fig:JointPlotRatio}Ratio of logs of estimated values of 
$p_{\alpha}(0)$ to logs of separability probabilities}
\end{figure}

Fig.~\ref{fig:JointPlot} is a joint plot of the logs of the $\alpha$-specific separability 
probabilities together with the logs of the estimated values (based on 10,100 moments) of the $y$-intercepts
$p_{\alpha}(0)$ for the 210 values $\alpha=\frac{1}{2},1,\frac{3}{2},2,\ldots 105$. The correlation between the pair of 210 log-values is almost perfect, that is, 0.99988 (cf. Fig.~\ref{fig:JointPlotRatio}).
\section{Fisher Information Equality Conjecture}
\begin{figure}
\includegraphics{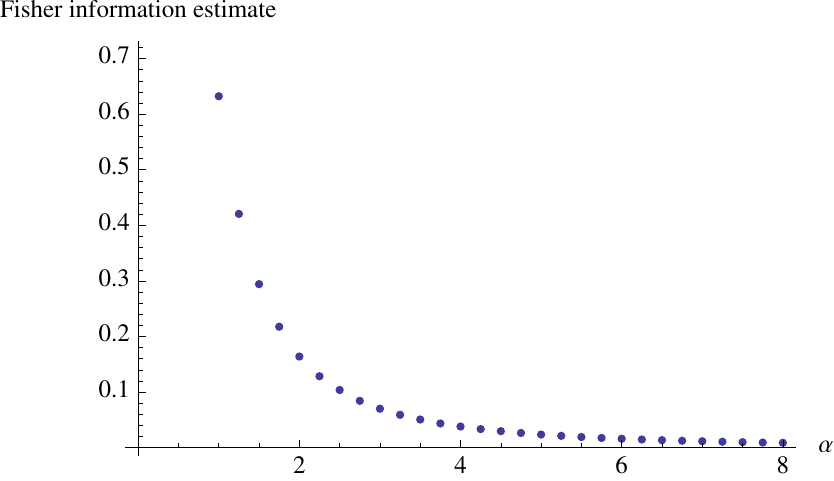}
\caption{\label{fig:FisherInformation}Estimates--based on the first 245 Hilbert-Schmidt moments of  $|\rho^{PT}|$--of the Fisher information of the $\alpha$-parameterized family of probability-distributions $p_{\alpha}(|\rho^{PT}|)$}
\end{figure}

\begin{figure}
\includegraphics{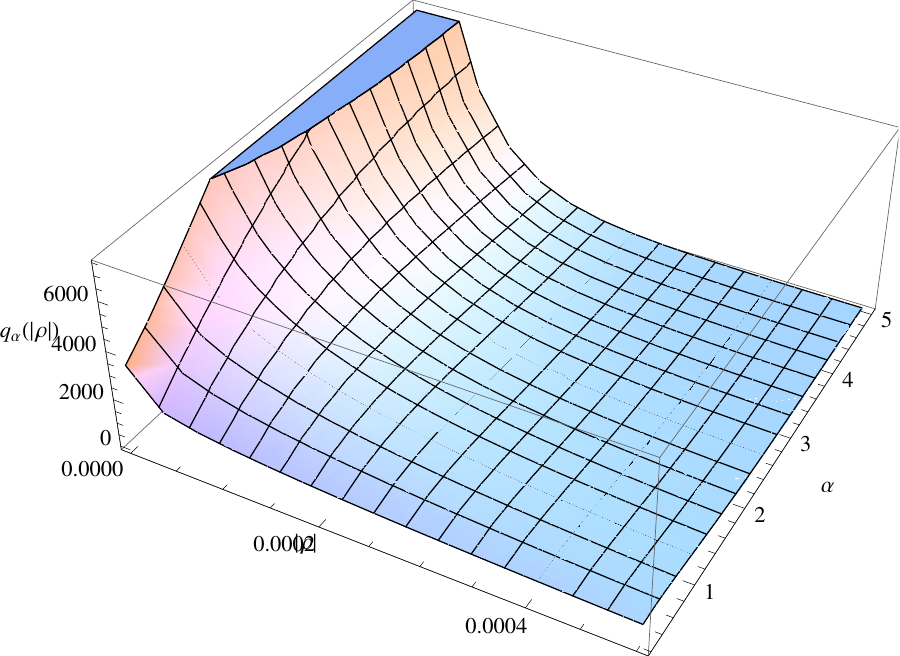}
\caption{\label{fig:Cappellini}The $\alpha$-parameterized family of probability distributions $q_{\alpha}(|\rho|)$ over $|\rho| \in [0,\frac{1}{256}]$ estimated on the basis of the first 100 Hilbert-Schmidt $\alpha$-specific moments  of $|\rho|$}
\end{figure}

\begin{figure}
\includegraphics{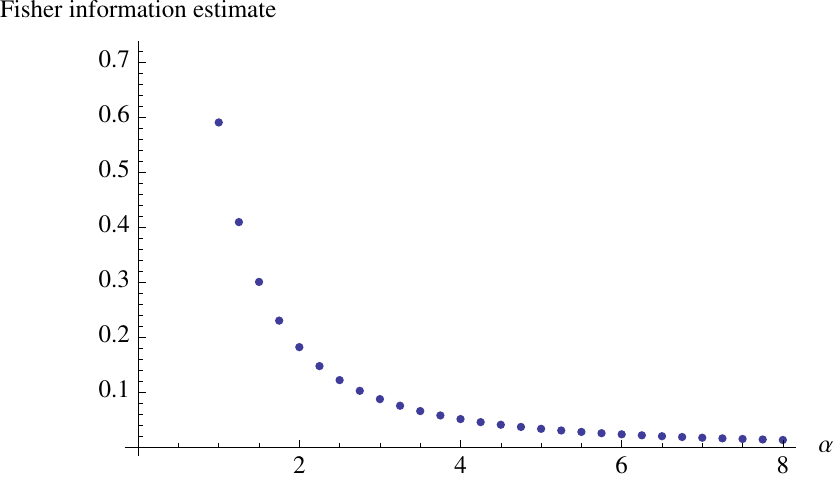}
\caption{\label{fig:FisherInformation2}Estimates--based on the first 100 Hilbert-Schmidt moments of  $|\rho|$--of the Fisher information of the 
$\alpha$-parameterized family of probability-distributions 
$q_{\alpha}(|\rho|)$. The correlation with the results in 
Fig.~\ref{fig:FisherInformation} is high ($> 0.99$).}
\end{figure}

The probability distributions $p_{\alpha}(|\rho^{PT}|)$, discussed above, form a one-parameter family--parameterized, obviously, by $\alpha$. Therefore,
we can inquire as to the Fisher information \cite{Frieden} of this family.
In Fig.~\ref{fig:FisherInformation}, we plot our estimates of this quantity  using the Legendre-polynomial probability distribution reconstruction 
procedure \cite{Provost} based on the first 245 Hilbert-Schmidt $\alpha$-specific moments (\ref{nequalzero}) 
of $|\rho^{PT}|$. One can, thus, conceive of a scenario in which one obtains a stream of determinants of random (with respect to Hilbert-Schmidt measure) partially transposed $4 \times 4$ density matrices 
($|\rho^{PT}|$). Then, employing the Fisher information (Fig.~\ref{fig:FisherInformation}) as a {\it prior} distribution over the Dyson-index-like parameter $\alpha$, one can derive--making use of Bayes' Theorem in conjunction with formulas for 
$p_{\alpha}(|\rho^{PT}|)$--a {\it posterior} distribution over $\alpha$ 
(cf. \cite{asher2}). 
(The extensive, diverse analyses reported in the treatise "Science from Fisher Information" \cite{Frieden} are concerned primarily with families of 
{\it translation-invariant} probability distributions--which clearly the family
$p_{\alpha}(|\rho^{PT}|)$ is not.)
In Fig.~\ref{fig:Cappellini} we show a plot comparable to 
Fig.~\ref{fig:AlphaFamilyPlot}, but now not employing the Hilbert-Schmidt moments of $|\rho^{PT}|$, given by (\ref{nequalzero}), but rather 
the Hilbert-Schmidt moments of $|\rho|$ \cite[sec. III, Fig. 3]{csz} 
(using Pochhammer symbol notation)
\begin{equation} \label{standardmoments}
\left\langle \left\vert \rho \right\vert ^{n}\right\rangle = 
\frac{256^{-n} (1)_n (a+1)_n (2 a+1)_n}{\left(3 a+\frac{5}{4}\right)_n
   \left(3 a+\frac{3}{2}\right)_n \left(3 a+\frac{7}{4}\right)_n},
\end{equation}
determining a one-parameter family of probability distributions 
$q_{\alpha}(|\rho|)$ over $|\rho| \in [0, \frac{1}{256}]$. (Dunkl has explicitly constructed the family $q_{\alpha}(|\rho^{PT}|$ \cite[Ex. 5.2]{CharlesDensities}.)
In Fig.~\ref{fig:FisherInformation2}, we show the counterpart to the Fisher  information plot 
Fig.~\ref{fig:FisherInformation}. Since these two plots appear to be quite comparable, it readily suggests the conjecture that the Fisher information is {\it identical} for the two families of probability distributions
$p_{\alpha}(|\rho^{PT}|)$ and $q_{\alpha}(|\rho|)$. (The correlation coefficient between the 32 values--$\alpha = \frac{1}{4},\frac{1}{2},\frac{3}{4},1,\ldots 8$--used in the two plots is remarkably high, 0.994524.) Since the "balanced" set of moments (\ref{BalancedMoments}) are in some sense "intermediate" between the moments of these two probability distributions 
((\ref{nequalzero}), (\ref{standardmoments})), one might also further speculate that
the associated Fisher information is identical to the hypothesized other two as well. 

If we turn from the Hilbert-Schmidt measure \cite{szHS,ingemarkarol} to the Bures measure \cite{szBures,ingemarkarol} over the generalized two-qubit states, then we have the set of moments (cf. \ref{standardmoments})
\begin{equation}\label{BuresMoments}
\left\langle \left\vert \rho \right\vert ^{n}\right\rangle_{Bures} =
 \frac{\Gamma (6 \alpha +2) 2^{-4 \alpha -8 n-1} \Gamma \left(n+\frac{1}{2}\right)
   \Gamma \left(n+\alpha +\frac{1}{2}\right) \Gamma (2 n+\alpha +1)}{\sqrt{\pi } \Gamma
   (n+2 \alpha +1) \Gamma (n+3 \alpha +1) \Gamma \left(2 n+3 \alpha
   +\frac{3}{2}\right)}.
\end{equation}
Our (Mathematica) 
estimate of the Hilbert-Schmidt Fisher information for the family 
$q_{\alpha}(|\rho|)$ at
$\alpha =\frac{1}{2}$ was 1.6595, while, in an independent Maple calculation, C. Dunkl obtained 1.660 (cf. \cite{CharlesDensities}). Dunkl also estimated the Bures counterpart of this value to be 0.9609. Further, for $\alpha=1$, Dunkl obtained 0.32532 for the Bures information, while our Mathematica Hilbert-Schmidt result was 0.590666. So, it would certainly seem, as might have been anticipated, that the generalized 
($\alpha$-parameterized0 two-qubit Hilbert-Schmidt and Bures Fisher information are different in nature.

\begin{acknowledgments}
I would like to express appreciation to the Kavli Institute for Theoretical
Physics (KITP) for computational support in this research, and to Charles Dunkl for his notes on hypergeometric probability distributions.
\end{acknowledgments}

\bibliography{Fisher}

\end{document}